\documentclass[twocolumn,prd,showpacs,10pt]{revtex4}
\usepackage{graphicx}
\usepackage{epsfig}
\usepackage{amsmath, amssymb}
\def\AA{{\cal A}}

\def\be{\begin{equation}}
\def\ee{\end{equation}}
\def\bea{\begin{eqnarray}}
\def\eea{\end{eqnarray}}
\def\ba{\begin{array}}
\def\ea{\end{array}}
\newcommand{\lsim}{\,\raise 0.4ex\hbox{$<$}\kern -0.8em\lower 0.62ex\hbox{$\sim$}\,}
\newcommand{\gsim}{\,\raise 0.4ex\hbox{$>$}\kern -0.7em\lower 0.62ex\hbox{$\sim$}\,}

\def\de{\mathrm{DE}}
\newcommand{\ca}{{c_a^2}}
\newcommand{\cs}{{c_s^2}}

\newcommand{\dep}{\delta p}

\newcommand{\HH}{{\mathcal H}}

\newcommand{\cc}{{\delta_{_0}}}
\newcommand{\cd}{{Q_0}}
\newcommand{\qtot}{{Q_{\rm tot}}}
\newcommand{\lar}{{\left(a\right)}}
\newcommand{\lkr}{{\left(k\right)}}
\newcommand{\lakr}{{\left(a,k\right)}}

\begin{document}

\title{Fingerprinting Dark Energy}
\date{October 12, 2009}

\author{Domenico Sapone}
\email{Domenico.Sapone@unige.ch}
\affiliation{D\'epartement de
Physique Th\'eorique, Universit\'e de
Gen\`eve, 24 quai Ernest Ansermet, CH--1211 Gen\`eve 4, Switzerland}
\author{Martin Kunz}
\email{M.Kunz@sussex.ac.uk}
\affiliation{Astronomy Centre, University of Sussex, Falmer, Brighton
BN1 9QH, UK}

\begin{abstract}
Dark energy perturbations are normally either neglected or else included in a
purely numerical way, obscuring their dependence on underlying parameters like
the equation of state or the sound speed. However, while many different 
explanations for the dark energy can have
the same equation of state, they usually differ in their perturbations so that
these provide a fingerprint for distinguishing between different models with
the same equation of state. In this paper we derive simple yet
accurate approximations that are able to characterize a specific class of models
(encompassing most scalar-field models) which is often generically called 
``dark energy''. We then use the approximate solutions to look at the impact
of the dark energy perturbations on the dark matter power spectrum and on the
integrated Sachs-Wolfe effect in the cosmic microwave background radiation.
\end{abstract}

\maketitle

\section{Introduction}

More than ten years after the supernova observations \cite{sn1,sn2} led to the general
acceptance that the expansion of the Universe is accelerating, we are still
far from a consensus on what is responsible for the acceleration. Although
there is a name for the phenomenon, dark energy, there is as of yet no
convincing physical explanation.

For this reason it is imperative to learn as much as possible from observations.
The fundamental observable considered so far is the equation of state parameter
$w$ which connects the average energy density $\rho(t)$ and the average pressure
$p(t)$ through $p(t)=w(t) \rho(t)$. This parameter characterizes the background
expansion rate and the distances. While some models make specific predictions,
like $w=-1$ for a cosmological constant, there are whole families of models that
can lead to any desired evolution of $w(t)$ (possibly with some weak constraints
like $w\geq-1$). Typical examples include scalar-field models like Quintessence
or K-essence, or generalized gravity models like scalar-tensor and $f(R)$ theories. These
theories cannot be ruled out based on a measurement of $w$ alone. However,
in general their perturbations evolve differently\cite{mfb,abp,lss,km,ks2,hs,aks}. 
The dark energy perturbations 
are therefore like a fingerprint of these models and, if measured, will allow to
discriminate much more precisely between competing theories, and hopefully will
allow to shed some light on the physical nature of whatever accelerates the
expansion of the Universe.

For many models, the behavior of perturbations at the linear level can be
described in terms of those of a fluid with a certain sound speed. This is
the case for Quintessence (canonical scalar-field) models for which the
sound speed is $\cs=1$, and for many K-essence models where the sound speed
is arbitrary. In this paper we concentrate on this class of models, and
assume in addition that both $\cs$ and $w$ are constant. This latter assumption
is usually violated, but the results should nonetheless allow insight
into the behavior of the perturbations. For models where the quantities
vary only slowly with time, we expect the results to still hold in an averaged
sense, due to the indirect nature of most observations. The big advantage of
making these assumptions is that it allows us to solve the perturbations
analytically under the additional condition of matter domination, leading to 
surprisingly simple results. In addition, when expressed in terms of the change 
of the gravitational potential relative to the case without dark energy 
perturbations then the simple formulae turn out to be a surprisingly good approximation
until today. Our results should be seen in this context, as it is of course
always possible to solve the perturbation equations numerically. However,
analytical results allow a much better insight and also an easy way to see
how the behavior changes as a function of the parameters.

Experiments tell us that the dark energy is at most very weakly coupled to the
things that we can observe directly, like galaxies and the cosmic microwave
background (CMB). It is therefore not only necessary to determine the dark energy
perturbations but also to connect them to actual observables. We use our simple analytical
results and make a small step in that direction, trying to establish the
impact of the dark energy perturbations onto the matter power spectrum
and the integrated Sachs-Wolfe (ISW) effect in the CMB. Our treatment here is 
far from complete and only an initial attempt to deal with the observational
impact of the dark energy perturbations. In a follow-up
publication \cite{ska} we will pay particular
attention to the question whether it is possible to observe the class of 
perturbations that we study here, using probes of lensing and of the matter
power spectrum.

In detail, the paper is organized as follows. We begin with a short discussion of
the perturbation equations to set the scene and to define our variables, 
and remind the reader of the solution for the matter
perturbations during matter domination. We then 
derive simplified solutions for the dark energy perturbations during
matter domination. Those expressions are a good fit in their respective
domain of validity, but once the dark energy starts to dominate they
deviate from the numerical solution. We then show that the function
$Q(k,t)$ used in \cite{aks} is well described by our solutions,
even after matter domination ends. This is the main result of the
paper. We finally use this observation to consider the impact of the
dark energy perturbations on the matter power spectrum and the growth
rate of the matter perturbations, as a function of $w$ and $\cs$. We
also investigate which aspect of the dark energy perturbations affects
the ISW effect most strongly, before concluding. The appendices give more details
on how we compared our analytical formulae to numerical results and
on the evolution of the decaying modes that were neglected.

% -----------------------------------------------------------------------

\section{First order perturbations in matter and dark energy}

Throughout this paper, we will use overdots to denote derivatives with respect
to conformal time $\tau$, related to cosmic time by $dt=a d\tau$. 
We will denote the physical Hubble parameter with $H$ and with $\HH$
the conformal Hubble parameter. We consider only spatially flat universes, and
our metric convention is defined by the line element,
\be
ds^{2} = a^{2} \left[ -\left( 1+2\psi \right) d\tau^{2} + \left( 1-2\phi\right) dx_{i}dx^{i} \right] .
\label{pert_newton_ds}
\ee
We are therefore working in the Newtonian or longitudinal gauge, which
influences the resulting perturbations especially on scales larger than
the Hubble horizon, $k\lsim aH$. On much smaller scales the choice of
gauge is less important, and observables are independent of the gauge
choice.

The perturbation equations for a fluid with equation of state parameter
$w=p/\rho$ are \cite{mabe,rd}
\bea
\delta' &=& 3(1+w) \phi' - \frac{V}{Ha^2} - 3 \frac{1}{a}\left(\frac{\dep}{\rho}-w \delta \right) \label{delta} \\
V' &=& -(1-3w) \frac{V}{a}+ \frac{k^2}{H a^2} \frac{\dep}{\rho}+(1+w) \frac{k^2}{Ha^2} \psi \\ \nonumber
&&-(1+w)\frac{k^2}{Ha^2}\sigma.  \label{v}
\eea
where $\delta = \delta\rho/\rho$ is the density contrast, 
$V=i k_j T_0^j /\rho$ is the scalar velocity perturbation (see also \cite{ks1}) and the prime means 
the derivative with respect to the scale factor $a$.
In this paper we will look only at fluids with vanishing anisotropic
stress, $\sigma=0$, so that there is a single gravitational potential, 
\be
k^2\phi =-4\pi Ga^2 \sum_j \rho_j \left(\delta_j+\frac{3aH}{k^2}V_j\right) 
\label{eq:phi}
\ee
and $\psi=\phi$.
The sum on the right hand side runs over all fluids.
For our purposes, we will assume the presence of a matter fluid with
$w=\delta p = 0$ as well as a dark energy fluid, parametrised by a
constant $w$ and a sound speed $\cs$ which determines the pressure perturbation
through
\be
\delta p = \cs \rho\delta+\frac{3aH\left(\cs-\ca\right)}{k^2}\rho V .
\ee
where $\ca\equiv \dot{p}/\dot{\rho}$ is called the adiabatic sound speed of the fluid.
This is not always a good parametrization, for example for models that
cross the phantom barrier $w=-1$ \cite{ks1} or when mimicking modified
gravity models \cite{ks2} (in which case also generically $\sigma\neq0$).
However, it covers a wide class of models, for example canonical scalar
fields (Quintessence, in which case $\cs=1$) and other scalar-field models 
(like K-essence, which allows for $\cs\neq1$).
As we only consider models with constant $w$, we have that
$\ca=w-\frac{\dot{w}}{3H\left(1+w\right)}=w$. The perturbation equations
(\ref{delta}) and (\ref{v}) become in this case
\bea
\delta' &=&  - \frac{V}{Ha^2}\left(1+\frac{9a^2 H^2\left(\cs-w\right)}{k^2}\right) \nonumber \\
&&-\frac{3}{a}\left(\cs-w\right)\delta+3\left(1+w\right)\phi' \label{deltap} \\
V' &=& -\left(1-3\cs\right) \frac{V}{a}+ \frac{k^2 \cs}{H a^2}\delta+(1+w) \frac{k^2}{Ha^2} \phi  \label{vp}
\eea

In general it is difficult to solve these equations. To simplify the problem,
we will assume that the universe is matter dominated. This means that the
Hubble expansion rate is given by
\be
H^2 = H_0^2 \Omega_m a^{-3} = \frac{8\pi G}{3} \rho_m
\ee
and that only the matter perturbations contribute to the gravitational
potential in Eq.~(\ref{eq:phi}). We can therefore first solve for the
matter perturbations alone, without considering the dark energy, and
then use the resulting gravitational potential as an external source
in the equations for the dark energy perturbations. Of course this
approximation, and even more the assumption of a matter dominated background evolution,
will change the results, and we will need to study what happens as
the assumptions break down.

The perturbation equations for the matter perturbations then are
\bea
\delta'_m &=&  - \frac{V_m}{Ha^2} +3\phi' \\
V'_m &=& - \frac{V_m}{a}+ \frac{k^2}{Ha^2} \phi \\
k^2 \phi &=& - \frac{3\Omega_m}{2a} \left(\delta_m+\frac{3aH}{k^2}V_m\right) 
\eea
As is well known, a solution to this set of equations is
\bea
\delta_m &=& \cc \left(a + 3 \frac{H_0^2 \Omega_m}{k^2}\right) 
= \cc a \left(1+3 \frac{H^2 a^2}{k^2} \right) \label{eq:deltam}\\
V_m &=& - \cc H_0 \sqrt{\Omega_m} a^{1/2}\\
k^2 \phi &=& -\frac{3}{2} \cc H_0^2 \Omega_m  \label{eq:phim}
\eea
where the constant $\cc$ sets the overall scale (since the equations
are linear).
This can be verified simply by inserting them into the differential
equations. The value of $\cc$ is set by the initial conditions and
is in general a function of $k$. Additionally the growth of the
matter perturbations is suppressed during radiation domination, so
that $\cc(k)$ is smaller than expected for the scales that enter
the horizon during radiation domination. However, the same happens
for the dark energy perturbations so that the effects cancel out
when comparing the perturbations in matter and dark energy. For
more details see Appendix \ref{app:camb} which discusses how we
compare the analytical results to CAMB \cite{camb}.

We see that the gravitational potential is constant as a function
of time on all scales. The matter perturbations grow linearly with $a$ 
on small scales, and they are constant on super-horizon scales 
(but their behavior on those large scales depends on the gauge choice).

% -----------------------------------------------------------------------

\section{Solutions for the dark energy perturbations during
matter domination}

We will now use the constant $k^2 \phi$ of the last section
to look for solutions to the general perturbation equations. We will
study them in different limits, and then compare the results with full
numerical solutions. See also \cite{ablr} for results using different approximations.

Generically, we expect at least three regimes with different behavior
of the perturbations:
\begin{itemize}
\item Perturbations larger than the causal horizon
\item Perturbations smaller than the causal horizon, but larger than
the sound horizon
\item Perturbations smaller than the sound horizon
\end{itemize}

\begin{figure}
\epsfig{figure=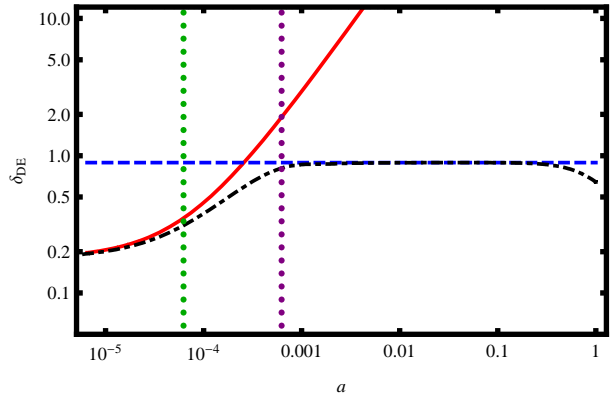,width=3.2in}
\caption{The figure shows the behavior of the variable $\delta_\de$. 
The black dot-dashed line is the numerical solution with $\cs=0.1$
and $w=-0.8$ for the mode $k=200 H_0$. 
The red solid line is the approximation on scales above the sound
horizon, Eq.~(\ref{eq:dsub}) and the blue dashed line is the approximation
to the scales below the sound horizon, Eq.~(\ref{eq:dpsh}). The two
vertical lines give the scale factor at which the mode enters the 
Hubble horizon (left line) and the sound horizon (right line). 
The numerical solution shows how the
perturbations decay at late times when matter domination ends, but radiation
was omitted from the numerical calculation to allow for a longer dynamic
range in $a$ to illustrate the different regimes.}
\label{fig:deltas}
\end{figure}

We start by looking at perturbations larger than the sound horizon,
$k \ll aH/c_s$. In this case, we neglect all 
terms containing the sound
speed in Eq.~(\ref{vp}), effectively setting $\cs=0$. The solution
for the velocity perturbation is (neglecting a decaying solution
$\propto 1/a$)
\be
V = - \cc (1+w) H_0 \sqrt{\Omega_m} a^{1/2} . \label{eq:vsub}
\ee
Up to the prefactor $(1+w)$ this is the same as for the matter velocity
perturbations. We find that this expression is valid on scales larger than
the sound horizon even if the sound speed is non-zero.

We can now insert this solution for the dark energy velocity perturbation
into Eq.~(\ref{deltap}). Again setting $\cs=0$ we find the solution
\be
\delta = \cc (1+w) \left( \frac{a}{1-3 w} + \frac{3 H_0^2 \Omega_m}{k^2} \right) \label{eq:dsub}
\ee
where we neglected a term proportional to $a^{3 w}$ which is decaying
as long as $w$ is negative. Not surprisingly, also this solution becomes
equal to the one for matter perturbations for $w\rightarrow 0$. Relative
to the matter perturbations the dark energy perturbations are suppressed
by the factor $(1+w)$. This factor is necessarily always there, as the
gravitational potential terms contain it. It can be thought of as
modulating the strength of the coupling of the dark energy perturbations
to the perturbations in the metric. For $w=-1$ the dark energy perturbations
are completely decoupled (in the sense that they do not feel metric
perturbations -- but they can still produce them if the dark energy perturbations
are not zero).

Just like Eq.~(\ref{eq:deltam}) the solution is composed of two terms,
where the second one dominates on scales larger than the horizon
($k<aH$) and the
first on smaller scales (but still larger than the sound horizon).
The super-horizon part of the solution is the same as for matter apart
from the overall $(1+w)$ factor, while the sub-horizon solution is
additionally suppressed by a factor of $1/(1-3w)$ relative to the
matter perturbations.

Although these factors can suppress the dark energy perturbations significantly
compared to the dark matter perturbations, especially if $w$ is close to $-1$,
the existence of a sound horizon is even more important. Inside the causal horizon,
the dark matter perturbations grow linearly with $a$ (until the perturbations
become non-linear). The dark energy perturbations on the other hand will
eventually encounter their sound horizon if $\cs>0$. Once inside the sound
horizon, they will stop growing. This means that the dark energy perturbation
spectrum is cut off on small scales.

To get a solution on small scales, $k\gg aH/c_s$, we start again with the equation
for the velocity perturbation. However, we expect the two terms with $k^2$
to cancel to a high degree to avoid large velocity perturbations, or in other
words
\be
\delta = -\frac{(1+w)\phi}{\cs} = 
\frac{3}{2} (1+w) \frac{H_0^2\Omega_m}{\cs k^2} \cc . \label{eq:dpsh}
\ee
We find that the dark energy perturbations stop growing and become
constant inside the sound horizon. 

The velocity perturbations are now given simply by using Eq.~(\ref{deltap})
and inserting Eq.~(\ref{eq:dpsh}):
\be
V = -3H a (\cs-w) \delta = - \frac{9}{2}(1+w)(\cs-w)\frac{H_0^3 \Omega_m^{3/2}}{\cs k^2} a^{-1/2} .
\label{eq:vsuper}
\ee
The extra term in brackets in Eq.~(\ref{deltap}) is not important for 
the scales of interest here.

Finally, we would like to remind the reader that these results have been
obtained under the assumption of a time-independent $w$ and $\cs$. On the
other hand, a $k$ dependence of $\cs$ is allowed.

As the horizons grow over time, a fixed wave number $k$ will correspond
to a scale that is larger than the causal horizon, $k<aH$, at early times,
and eventually it will enter the causal horizon and later the sound horizon.
This makes it possible to illustrate the behavior of the perturbations
in the different regimes in a single figure: In Fig.~\ref{fig:deltas} we
plot the numerical solution for the dark energy density contrast
for $k=200 H_0$ as well as the expressions (\ref{eq:dsub}) and (\ref{eq:dpsh}).
It is easy to see how the perturbations start to grow inside the causal
horizon but how the growth stops when the sound horizon is encountered
and pressure support counteracts the gravitational collapse.

% -----------------------------------------------------------------------

\section{Dark energy domination and $Q$}

Matter domination was a crucial ingredient to compute the behavior
of the dark energy perturbations, since the gravitational potential
$\phi$ is constant while matter dominates the expansion rate and
the total perturbations. However, dark energy comes to dominate
eventually, and then the potential starts to decay and the perturbations
grow more slowly or start to decrease. This is also visible in
Fig.~\ref{fig:deltas} at very late times where the numerical solution
for $\delta$ starts to decrease.

It is difficult to capture this behavior accurately. But in \cite{aks}
we introduced the variable $Q(k,a)$ to describe the change of the gravitational
potential due to the dark energy perturbations. $Q$ is defined through
\be
k^2\phi =-4\pi Ga^2 Q \rho_m \left(\delta_m+\frac{3aH}{k^2}V_m\right) 
\label{eq:Q_def}
\ee
If the dark energy or modification of gravity does not contribute to
the gravitational potential (for example if the dark energy is a
cosmological constant) then $Q=1$. Otherwise $Q$ will deviate from
unity, and in general it is a function of both scale and time.

\begin{figure}
\epsfig{figure=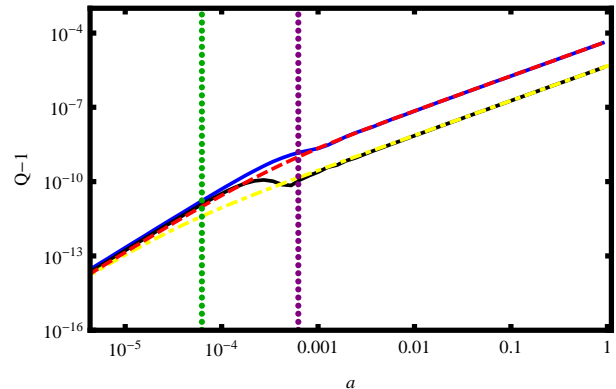, width=3.2in}
\caption{The figure shows the behavior of the variable $\qtot-1$ for modes above 
and below the sound horizon. The solid lines are CAMB output (blue / upper line: $\cs=0.1$,
black / lower line: $\cs=1$) and the dashed lines use Eq.~(\ref{eq:qtot}), 
red (upper) for $\cs=0.1$
and yellow (lower) for $\cs=1$. The vertical lines show the sound horizons, the left
line for $\cs=1$ and the right line for $\cs=0.1$.
The behavior on scales larger than the sound horizon
is the same in both cases, but the growth of the perturbations slows down as the
mode enters the sound horizon. As the modes enter the sound horizon earlier
for $\cs=1$ they stop growing earlier and stay smaller. We used again $w=-0.8$
and $k=200 H_0$.}
\label{fig:qtot}
\end{figure}

Introducing the comoving density perturbation 
$\Delta \equiv \delta + 3 aHV/k^2$, and looking at Eq.~(\ref{eq:phi})
we see that we can compute $Q$ with
\be
Q - 1 =  \frac{\rho_\de \Delta_\de}{\rho_m \Delta_m} . 
\ee
Just using the results during matter domination, we find that the resulting
expression for $Q$ is surprisingly accurate even at late times (see Fig.~\ref{fig:qtot}). 
The reason is that both fluids, dark energy and matter,
respond similarly to the change in the expansion rate so that most of the
deviations cancel. We find that the sub-soundhorizon expression below is
accurate at the percent level, while on larger scales there are deviations
of about 10 to 20\% by today (depending on $w$). The latter can be corrected
"by hand" in order to obtain a more precise formula, but the expressions are
sufficiently accurate for our purposes and we keep them as they are.

For matter, and during matter domination, the comoving density perturbation
is extremely simple, $\Delta_m = \cc a$. For the dark energy we find
\be
\Delta_\de = \cc \frac{1+w}{1-3w} a = \frac{1+w}{1-3w} \Delta_m
\ee
on scales larger than the sound horizon. This means that the
relative strength of the comoving density perturbations in the dark energy
and the dark matter is constant on large scales, with those in the
dark energy being sub-dominant for $w<0$ --- for $w$ close to $-1$
the prefactor is approximately $(1+w)/4$. Using the scaling of the
energy density in matter and dark energy, we can derive that
\be
Q-1 = \left( \frac{1-\Omega_m}{\Omega_m} \right)
      \left( \frac{1+w}{1-3w} \right) a^{-3w}
    \equiv \cd a^{-3 w} . \label{eq:qm1}
\ee
Here we defined a constant $\cd$ since this expression will appear
frequently later on.
For $\Omega_m=0.25$ and $a=1$ $\cd$ interpolates smoothly between 
$0$ for $w=-1$ and $3$ for $w=0$. For $w=-0.8$ we have 
$Q(a=1)-1 =\cd \approx 0.18$, that means that we do not expect more
than about a 20\% deviation of $Q$ from $1$ even on large scales,
given current observational limits on $w$. We will keep using
$w=-0.8$ to illustrate what we can maximally expect to observe.

On the other hand,
on small scales the growth of dark energy perturbations is stopped
by pressure support. The dark energy perturbations stop growing once
they are inside the sound horizon, while the matter perturbations
(for which there is no sound horizon) continue to grow. This leads to
\be
\Delta_\de \approx \frac{3}{2} (1+w) \left(\frac{Ha}{c_s k}\right)^2 \Delta_m
\propto \Delta_m/a
\ee
on scales below the sound horizon. Here we neglected the dark energy
velocity perturbations, since their contribution to $\Delta_\de$ 
is suppressed 
by a factor proportional to $(Ha/k)^2$ relative to $\delta_\de$.
From this expression, we then find that
\bea
Q-1 &=& \left( \frac{1-\Omega_m}{\Omega_m} \right)
      \frac{3}{2} (1+w) \left(\frac{Ha}{c_s k} \right)^2 a^{-3w} \\
    &=& (1-\Omega_m) \frac{3}{2} (1+w) \frac{H_0^2}{\cs k^2}
       a^{-1-3w}\label{eq:qm2} ,
\eea
where the result is more accurate when using the matter $Ha$ as done above.
Also on small scales, deviation of $Q$ from $1$ is maximal at late times
for any dark energy that leads to acceleration, $w<-1/3$. But in general
it will be suppressed relative to the large scales by the lack of growth
of the dark energy perturbations once they are inside the sound horizon.

For the variable $Q$ we can construct a unified formula which accounts 
both for modes below and above the sound horizon,
\be
\qtot-1 = \frac{1-\Omega_{m}}{\Omega_{m}}\left(1+w\right)\frac{a^{-3w}}{1-3w+\frac{2k^{2}\cs a}{3H_0^{2}\Omega_m}}\label{eq:qtot}
\ee

In Fig.~\ref{fig:qtot} we compare $\qtot$ 
with the numerical solution from CAMB for different values of the dark energy 
sound speed. The formula for $\qtot$ interpolates between the two asymptotic
regions. Close to the sound horizon it is not very accurate, but it is sufficient
to use in e.g. Fisher-matrix codes to estimate the size of effects due to the
presence of dark energy perturbations.

% -----------------------------------------------------------------------

\section{Impact on some observational quantities}

\subsection{Growth and shape of the matter power spectrum} % growth and shape

What impact does the change in $\phi$ from $Q\neq1$ have on the matter 
perturbations? Notice that the effect of the change in $H$ from the onset
of dark energy domination at late times is expected to be larger than the
the effect from the presence or absence of dark energy perturbations, but
here we want to quantify only the latter.
We can write $\phi = \phi_m + \phi_\de$ where $\phi_m$ is the solution
given by Eq.~(\ref{eq:phim}) and $\phi_\de$ is $(Q-1)\phi_m$. Since the
differential equation is linear, the solution with the total source is
just the sum of the solution for each source (plus, as before, a decaying
solution $\propto 1/a$). The expressions we have found for $Q$ are always
power-laws in $a$, which guarantees a simple form for the velocity perturbation.

Even though the dark matter does not have a sound horizon, the
solutions will now depend on whether the $k$ mode in question is
larger or smaller than the sound horizon of the dark energy,
simply because the deviation of $\phi$ from $\phi_m$ depends
on this. For $k$-modes larger than the dark energy sound horizon,
we find for the matter velocity perturbation
\be
V_m = - \cc H_0 \sqrt{\Omega_m} \sqrt{a}
     \left\{ 1+ 
      \left( \frac{\cd}{1-2w} \right)
      a^{-3 w} \right\}
\ee
where $w$ is the equation of state parameter of the dark energy
fluid, and the constant $\cd$ was defined in Eq.~(\ref{eq:qm1}). 
At late times $V_m$ will deviate from this formula because of
the change in the expansion rate, but we are again interested in the
impact of the perturbations, corresponding to the factor in curly
brackets. The factor is less sensitive that the velocity itself,
and it tracks the numerical result closely until dark energy domination
sets in, giving at least the right order of magnitude even today.
For the optimistic case with $w=-0.8$ we find a deviation
of about 6.5\% at late times due to the presence of the dark energy
perturbations. The fully numerical integration for this case
gives about 4.5\% change. The agreement is not as good as for
the dark energy perturbations themselves, but still acceptable, especially
since we are dealing with an indirect effect.

This solution we can insert in the one for the density
perturbation. In this case, $\phi' \neq 0$, but since
we are again dealing with a linear equation, we end up with
a sum of three solutions, one for each source (the two parts of
$V_m$ and $\phi'$) as well as the constant peculiar solution:
\bea
\delta_m &=& \cc \biggl\{ 3 \frac{H_0^2 \Omega_m}{k^2} 
\left(1+\frac{3}{2} \cd a^{-3w}\right) \label{eq:dm_mod}\\
&& + a \left(1+ \frac{\cd a^{-3 w}}{1-5 w+6w^2} \right) \biggr\} \nonumber
\eea
On large scales, larger than the causal horizon, the matter perturbations 
are therefore enhanced by a factor $1+(3\cd)/2 a^{-3w}$, which is of the
same order as $Q$. On smaller scales, but still larger than the sound
horizon, the factor is $1+\cd a^{-3w}/(1-5w+6w^2)$ which is smaller by a
factor of 10 or so (depending on $w$). Inside the sound horizon, the
growth of the dark energy perturbations is suppressed.

\begin{figure}
\epsfig{figure=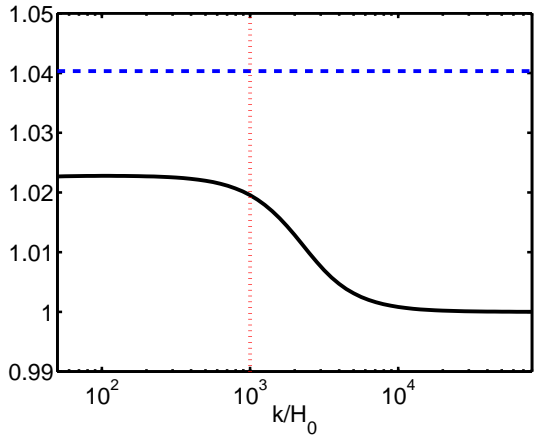, width=3.2in}
\caption{The modification of the matter power spectrum $P(k)$: The solid
black line shows how the matter power spectrum is enhanced outside the
sound horizon (vertical red dotted line). The blue dashed line shows
the prediction from Eq.~(\ref{eq:dm_mod}) for the modification outside
the sound horizon. Here we used $w=-0.8$ and $\cs=10^{-6}$.}
\label{fig:pkmod}
\end{figure}

It is customary to consider two different aspects of the matter 
perturbations: the power spectrum today and the growth
rate of the perturbations. For the power spectrum today we can
for example look at the ratio of $\delta_m^2$ with dark energy perturbations
to $\delta_m^2$ without the perturbations. For our usual optimistic
benchmark with $w=-0.8$ we expect a 4\% enhancement of $P(k)$ on scales
larger than the sound horizon. A numerical calculation shows that it
is closer to 2\%, see Fig.~\ref{fig:pkmod}.

Observationally there are two difficulties to detect the effect: first,
it is unclear what happens with the dark energy perturbations on scales
where the matter perturbations become non-linear, especially if those
scales are larger than the dark energy sound horizon. Second, for a
dark energy model with a very large sound speed like Quintessence, for
which $\cs=1$, the matter power spectrum will only be affected on the
largest scales which have not been observed. It is obviously much easier
to measure the sound speed if it is low. However, the feature visible
in Fig.~\ref{fig:pkmod} would provide a clear signature for the presence
(and size) of a sound-horizon, especially if the corresponding change
in the matter velocity field could be detected as well.

A potentially more promising place to look for the impact of the dark
energy perturbations is in way the matter perturbations grow over time.
Often the impact of the dark energy on the growth rate of the matter
perturbations is parametrised in terms of a growth index $\gamma$ \cite{lahav,wast}, 
defined through
\be
\frac{d \log(\Delta_m)}{d\log(a)} = \Omega(a)^\gamma
\label{eq:growthdef}
\ee
We will now try to connect our expression (\ref{eq:dm_mod}) to $\gamma$ through some rather
crude approximations which should nonetheless be good enough to result
in an order of magnitude estimate of the change of $\gamma$ due to the
presence of the extra perturbations.
On small scales (but always larger than the dark energy sound horizon) the 
differences between $\delta_m$ and $\Delta_m$ are
suppressed by $(Ha/k)^2$ and we have
\be
\Delta_m(k\gg Ha) \approx \cc a \left( 1+\cd \frac{a^{-3w}}{1-5w+6w^2} \right)
\ee
Using $\Delta_m^{(0)}=\cc a$ to model partially the late-time change of the
expansion rate, and assuming that this term obeys the form
of Eq.~(\ref{eq:growthdef}) with unperturbed growth index $\gamma_0$ we
find by performing the derivative
\bea
\frac{d \log(\Delta_m)}{d\log(a)} &=& \frac{\Delta_m^{(0)'} a}{\Delta_m^{(0)}}
   - \frac{3 w \cd a^{-3w}}{\cd a^{-3w} + (1-5 w + 6 w^2)} \nonumber \\
   &=& \Omega(a)^{\gamma_0} + \frac{3(Q-1)}{5-6w-Q/w} \label{eq:growthmod}
\eea
where we used Eq.~(\ref{eq:qm1}) to reintroduce $Q-1$. In general this
cannot be cast in the form of $\Omega_m(a)^\gamma$ since the dark energy
perturbations change the matter growth rate even during matter domination
where $\Omega_m(a)=1$. We can however connect to that form at least in
the limit when $\Omega_m(a)$ just starts to deviate from unity. 

We would like to end up with $\Omega(a)^{\gamma_1}$ with
$\gamma_1=\gamma_0+\epsilon$. If $\epsilon$ is small enough then we
can use that $\Omega_m(a)^\epsilon\approx 1+\epsilon \log(\Omega_m(a))$
and if we are close to $\Omega(a)=1$ then additionally
$\log(\Omega_m(a))\approx\Omega_m(a)-1$, giving finally
\be
\epsilon(\Omega_m(a)-1)\approx \frac{3(Q-1)}{5-6w-Q/w}
\label{eq:oureps}
\ee

According to \cite{lica} (see also \cite{bari}), 
the growth index depends on $Q$ through the
combination
\bea
\gamma &=& \frac{3(1-w-A(Q))}{5-6w}  \label{eq:lica1} \\
A(Q)   &=& \frac{Q-1}{1-\Omega_m(a)} \label{eq:lica2} .
\eea
so that in terms of our $\epsilon$ above
\be
\epsilon = \frac{-3(Q-1)}{(1-\Omega_m(a))(5-6w)}
\ee
which is sufficiently close to Eq.~(\ref{eq:oureps}), given the crudeness
of the approximations used. Numerically the expression above is very close
to Eq.~(\ref{eq:growthmod}) -- closer than to the full numerical solution
which remains more constant at late times. Both predict the correct deviation
at early times, though.

The denominator of the function $A$ is
\be
1-\Omega_m(a) = \frac{(1-\Omega_m) a^{-3w}}{\Omega_m+(1-\Omega_m) a^{-3w}} .
\ee
Not surprisingly, this is $1-\Omega_m$ today, so that $A\approx(1+w)$,
while at early times
the second term in the denominator is suppressed so that for $a\ll 1$
$A$ becomes $\Delta_\de/\Delta_m\approx (1+w)/4$ on large scales. This
has to be compared to $1-w\approx 2$. The dark energy perturbations
thus change the growth index by a few percent or about $0.02$ 
for $w=-0.8$ on scales larger than the sound horizon. 
This will be challenging to measure even by full-sky surveys like the
proposed Euclid satellite mission which expects to achieve an error
on $\gamma$ of less than this \cite{aks} but only if enough modes can
be measured, i.e. if the sound speed is close
to zero. Once the perturbations enter the sound horizon, they stop to grow and
so their impact on the matter perturbations decreases and becomes
rapidly negligible and thus impossible to detect. However, notice that this is purely
the change of the growth rate due to the {\em perturbations} in the dark 
energy fluid! We defer a more detailed investigation of the detectability
of the dark energy perturbations in the dark matter power spectrum
to a later publication \cite{ska}.

\subsection{The integrated Sachs-Wolfe effect}

After the matter power spectrum, we focus on the integrated Sachs-Wolfe (ISW)
effect for cosmological models with non-zero contribution from the dark energy
perturbations. The motivation
is that the ISW part of the spectrum is the most affected by the dark 
energy, see \cite{wellew,beandore,bms,ddw}.

The ISW effect results from the late-time decay of gravitational 
potentials. The total blueshifting or redshifting of the CMB photons caused by the 
change in the potential during the passage of the photons
induces temperature fluctuations \cite{Sachs-Wolfe}:
\be
\zeta = \frac{\Delta T\left(\hat{n}\right)}{T_{0}}=2\int{\frac{\partial\phi}{\partial\tau}}d\tau= -2\int_{0}^{\chi_{_H}}{a^{2}H\frac{\partial\phi}{\partial a}}d\chi,
\ee
where $\tau$ denotes conformal time. In the last step, we have replaced the integration variable by the comoving distance $\chi$ which is related to the conformal time by $d\chi =-cd\tau=-cdt/a$; here we assume a zero anisotropic stress component for all the species in the Universe. 

In Fourier space, the derivative of the gravitational potential with respect to the scale factor $a$ can be expressed as:
\bea
&&\phi'=-\frac{3}{2} \frac{H_{0}^{2}\Omega_{m}}{ak^2} \left\{Q\left(a,k\right)\Delta'_{m}\left(a,k\right) + \right. \nonumber \\
&&+\left. Q'\left(a,k\right)\Delta_{m}\left(a,k\right)-\frac{1}{a}Q\left(a,k\right)\Delta_{m}\left(a,k\right)\right\}. \label{psidot}
\eea
where the prime denotes the derivative with respect the scale factor.
There are three main ways in which the dark energy perturbations
change the ISW effect: They change $\Delta_m$ itself, as discussed
in the last section, both the shape (a very small effect) and the
growth. They change $\phi'$ additionally through the presence of $Q$
in the last and first term of Eq.~(\ref{psidot}) and their variation
enters as well through $Q'$.

In linear perturbation theory all $k$ modes evolve independently, so
that we can decompose the dark matter density contrast as
\be
\Delta_{m}\lakr=aG\lakr\Delta_{m}\lkr .
\ee
Here $\Delta_{m}\lkr\equiv \Delta_m(a=1,k)$.
The so-called growth factor $G$ is usually independent of scale: during
matter domination $G$ is constant and the late-time change in the
expansion rate affects all scales equally. However, the contribution from 
the dark energy perturbations induces a scale dependence because of
the existence of the dark energy sound horizon, see for instance \cite{ddp}. For the growth rate,
we will use the $\gamma(a,k)$ from Eqs.~(\ref{eq:lica1}) and (\ref{eq:lica2}),
which provides a sufficiently good approximation. 

We can then write Eq.~(\ref{psidot}) as:
\be
\phi' = -\frac{3}{2}\frac{H_{0}^{2}\Omega_{m}}{k^2}\frac{\partial}{\partial a}\Big \{G\lakr Q\lakr \Big\}\Delta_{m}\lkr . \label{eq:psiprimeG}
\ee

The line of sight integral for the ISW-temperature perturbation $\zeta$
can now be written as
\be
\zeta = \int_{0}^{\chi_{_H}}{d\chi W_{\zeta}\left(\chi\right)\Delta_{m}\lkr}
\ee
where we introduced the weight function:
\be
W_{\zeta}\left(\chi\right)=\frac{3}{c^3}\frac{H_{0}^{2}\Omega_{m}}{k^{2}}a^{2}H\frac{\partial}{\partial a}\Big \{G\lakr Q\lakr \Big\}
\label{windowISW}
\ee
which allows the expressions for the ISW-auto spectrum $C_{\zeta\zeta}\left(\ell\right)$ to be written in a compact notation, applying the Limber-projection \cite{Limber} in the flat-sky approximation, for simplicity:
\be
C_{\zeta\zeta}\left(\ell\right)=\int_{0}^{\chi_{_H}}{d\chi\frac{W_{\zeta}^{2}\left(\chi\right)}{\chi^{2}}\bar{P}_{\Delta\Delta}\left(k =\ell/\chi\right)} .
\label{clstoplot}
\ee

$\bar{P}_{\Delta\Delta}\left(k\right)$ is the linear matter power spectrum today, which can be written as:
\be
\frac{k^3\bar{P}_{\Delta\Delta}\left(k\right)}{2\pi^2}=\delta^{2}_{H}\left(\frac{k}{H_0}\right)^{n+3}T^{2}\left(k\right) .
\ee
Here $\delta_{H}$ is the amplitude of the present-day density fluctuations at the Hubble scale 
and $T\left(k\right)$ is the transfer function for CDM. We adopt the fit by 
Eisenstein \& Hu \cite{ehu}. We have neglected the additional impact from the dark energy
perturbations onto the dark matter power spectrum as it is only of the order of a few
percent (see the discussion in the last section). The Eisenstein \& Hu fit agrees with
the CAMB output to a precision of about 4\% for both low and high values of $\cs$,
more than sufficient for the purposes of this section.

\begin{figure}
\epsfig{figure=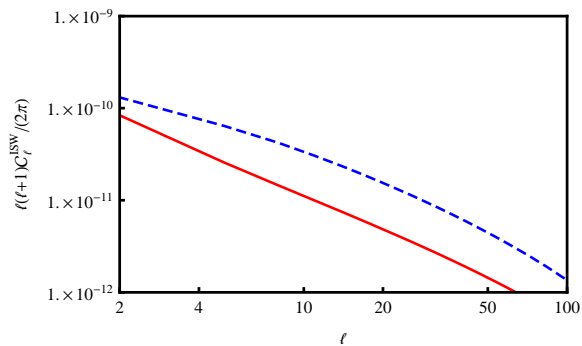,
  width=3.0in}
\caption{ISW power spectrum for $\cs= 10^{-4}$ (red solid line) and $\cs= 1$ 
(blue dashed line) for a model with $w=-0.8$.}
\label{fig:ISW-Cls}
\end{figure}

In Fig.~\ref{fig:ISW-Cls} we plot $\ell\left(\ell+1\right)C_{\ell}/(2\pi)$ using 
Eq.~(\ref{clstoplot}) for two different values of the dark energy sound speed. The 
differences between the two curves come from the term $(\partial (QG)/\partial a)^2$. 
Since the ISW power spectrum depends on the derivatives of the product of the growth 
factor $G\lakr$ and $Q\lakr$ we need to look at
\be
(GQ)' = G'Q + GQ' \label{eq:sigma}
\ee
The deviation of $Q$ from $1$ is never enough to explain the differences between the two
curves in Fig.~\ref{fig:ISW-Cls}. Indeed, taking $Q=1$ while keeping $Q'$ barely changes 
the results. The relative size between the first and second term depends 
mainly on $Q' G/G'$. From Eq.~(\ref{eq:qtot}) we see that $aQ'\propto (Q-1)$
with a proportionality factor of about 2 to 3. This is a small number, but it is boosted
by $G'\ll G$, since $aG'/G=\Omega_m(a)^\gamma -1$. The dark energy slows down the growth
of the dark matter perturbations so that $G'<0$. However, $Q'>0$ because of the relative
increase of the dark energy density enhancing the importance of the dark energy 
fluctuations. The two
contributions will partially cancel and so decrease the result, unless one term
dominates strongly, e.g. for modes inside the sound horizon. This is the reason 
why the ISW contribution to the CMB power
spectrum decreases as the sound speed increases.

We can illustrate the effect by looking at the 
{\em gravito}-power spectrum,
\be
P_{\phi'\phi'}= \left( \frac{3H_{0}^{2}\Omega_{m}}{2k^2}\right)^{2}\left\{\frac{\partial}{\partial a}G\lakr Q\lakr\right\}^{2}\bar{P}_{\Delta\Delta}\left(k\right) .
\label{eq:psipsi-G}
\ee
If we assume that the dark energy does not cluster (i.e. $Q=1$), Eq.~(\ref{eq:psipsi-G}) reads:
\be
P_{\phi'\phi'}= \left( \frac{3H_{0}^{2}\Omega_{m}}{2k^2}\right)^{2}\left\{\frac{\partial}{\partial a}G\lar\right\}^{2}\bar{P}_{\Delta\Delta}\left(k\right).
\label{phidotPS-noq}
\ee
and the growth factor becomes again a function of time only.

Comparing Eq.~(\ref{eq:psipsi-G}) and Eq.~(\ref{phidotPS-noq}) we can define a magnification parameter for the ISW power spectrum:
\be
\AA^2= \left\{\frac{d\left(G\lakr Q\lakr\right)/da}{dG\lar/da}\right\}^{2} .
\ee
This is precisely the expression discussed above, $\AA=Q+Q' G/G'$.

\begin{figure}
\epsfig{figure=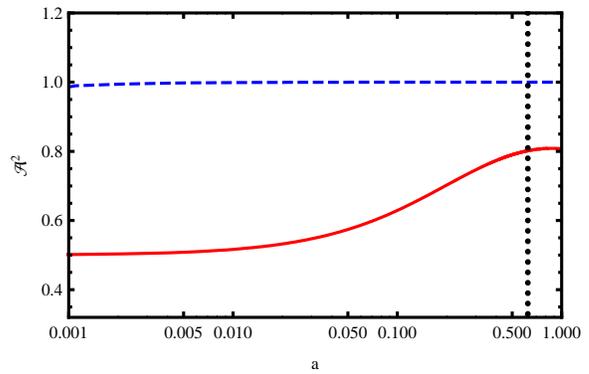,
  width=3.0in}
\caption{We plot the magnification factor $\AA^2$ with $k=200 H_0$ and $w=-0.8$ for two different values of the sound speed: $\cs= 10^{-4}$ (red solid line) and $\cs= 1$ (blue dashed line). The vertical line at $a=0.62$
shows when the $k$-mode enters the sound horizon for $\cs= 10^{-4}$ (for the high sound speed the mode
enters at $a=0.62\times10^{-4}$).}
\label{fig:magnification}
\end{figure}

In Fig.~\ref{fig:magnification} we plot the magnification factor for two values of $\cs$ for $k=200 H_0$.
For a sound speed equal to $1$ the dark energy perturbations enter the sound horizon very early and stay
small until today, even when taking into account that the dark energy density grows relative to the dark
matter density. As expected, they do not affect the ISW effect significantly in this case. For small sound
speeds the dark energy perturbations partially cancel the contribution from $G'$ and decrease $\AA$ by
about 30\% for scales above the dark energy sound horizon. Their slower growth inside the sound horizon
leads to a smaller cancellation, as can be seen in Fig.~\ref{fig:magnification}. 
Based on these observations, we expect that a low sound speed decreases the ISW effect (which goes 
like $\AA^2$) by about 50\%, consistent with Fig.~\ref{fig:ISW-Cls}. Overall, it seems that $Q'$
can provide a more sensitive probe of the dark energy perturbations than $Q$.

\section{Conclusions}

In this paper we have derived simple analytical expressions for the behavior
of the perturbations in fluid dark energy models with vanishing anisotropic
stress, equations (\ref{eq:vsub}) to (\ref{eq:vsuper}). These expressions are 
valid for models with constant $w\leq0$ and 
$\cs\ge0$. They were derived under the assumption of matter domination, but
they allow to compute the function $Q(k,a;w,\cs)$ which describes the deviation of
the Poisson equation from the case without dark energy perturbations and which
is relatively insensitive to the late-time deviations from matter domination. The
expressions for $Q$ are given in Eq.~(\ref{eq:qm1}) for scales larger than the
sound horizon and Eq.~(\ref{eq:qm2}) for scales smaller than the sound horizon.
We also give a single interpolating equation in Eq.~(\ref{eq:qtot}) which is
useful for Fisher-matrix calculations that include dark energy perturbations.  
In models without anisotropic stress $Q$ completely characterizes
the dark energy perturbations and represents a fingerprint that allows to
differentiate between different models with the same background expansion rate 
but a different evolution of the perturbations.

We expect our results to hold generically for scalar-field like models.
Large changes can appear because of rapidly varying (especially oscillating)
$w$ leading to resonance-like behavior (see e.g. the rather contrived 
"phaxion" model in \cite{ks1}), for models with non-zero effective
anisotropic stresses (like DGP \cite{dgp, ks2}) or couplings between
the dark energy and the dark matter \cite{couplings1,couplings2,dark_degen}, so
such models will lead to different results.

We find that the dark energy perturbations are always smaller than the
perturbations in the dark matter, at least by a factor $(1+w)$, but on
scales larger than the dark energy sound horizon they are only suppressed
by an additional factor of order unity. The dark energy perturbations do
not grow on scales smaller than the sound horizon, so that these 
perturbations are much more
suppressed relative to the dark matter. The impact of the dark energy
perturbations on the gravitational potential is additionally influenced
by the relative energy density, so that measurable deviations tend to
appear only at late times.

To demonstrate the usefulness of the equations we then used the formula 
for $Q$ to investigate the change in the dark
matter power spectrum and the ISW effect if $w=-0.8$. The changes in the matter
power spectrum at late times are of the order $(Q(a,k)-1)/5$ which
corresponds to a few percent on scales larger than the sound horizon
of the dark energy. The growth index $\gamma$ is changed by about
0.02 on the same scales. We further find that the impact on the ISW effect 
is due to the growth of the dark energy perturbations, $Q'$ having a larger
effect on $\phi'$ than naively expected from the size of $Q$. 
It is much larger than the impact of the dark energy perturbations
on the matter power spectrum, 
but because of cosmic variance it is more difficult to constrain
observationally.

The dark energy perturbations of the class of models investigated here
vanish as $w\rightarrow -1$. However, if $w$ is different from $-1$ and
especially if the sound speed of the dark energy is less than the speed
of light, then there is hope that the effects from the dark energy
perturbations could be seen with cosmological observations. Although
very challenging, it is nonetheless worth the effort since the
perturbations provide a much more precise ``fingerprint'' of the dark
energy than the equation of state parameter $w$.

\begin{acknowledgments}

D.S. is supported by the Swiss NSF, M.K. is supported by STFC (UK).

\end{acknowledgments}

\begin{appendix}

\section{Comparing with CAMB \label{app:camb}}

Since we are dealing with linear perturbation theory, all $k$-modes evolve independently
and for each $k$-mode the perturbations depend linearly on the normalization given by
the constant $\cc$. As we can choose this constant arbitrarily for each $k$-mode, we
should really think of it as $\cc(k)$. In the standard cosmology, its value is set in the
very early universe by inflation. This scenario then provides the initial conditions
for the differential equations. However, when we compare our results with numerical
solutions from e.g. CAMB, we have to take into account as well that the Universe was
radiation dominated at early times. Since both the expansion rate of the Universe and
the dominant contribution to the gravitational potentials are different during radiation
domination, we expect to find a different behavior for the matter and dark energy
perturbations. It turns out that
the change in the expansion rate strongly suppresses the growth rate of the perturbations,
leading to an only logarithmically growing solution \cite{Paddy}.

This is not directly relevant to our solutions as we are content to limit our expressions
to the matter dominated and later epochs. One nuisance is that we have to disregard
radiation
domination in some of the figures where we prefer a larger dynamical range to show the
(formal) evolution of the perturbations in the different regimes. We note this in the
figure captions where applicable. Another issue concerns the normalization
for comparison with numerical codes: as the perturbation growth of sub-horizon modes is
delayed by the radiation dominated epoch, the $k$ modes which enter the horizon during
that period end up with a lower normalization than expected if they are normalised at
early times. The pragmatic solution here is to normalize the
perturbations instead in the late universe, after the onset of matter domination,
\be
\delta_m\left(a=a_1\right)=\delta_{\rm
in}=\cc\left(a_1+\frac{3H_{0}^{2}\Omega_{m_0}}{k^{2}}\right)
\ee
which fixes $\cc(k)$ in terms of $\delta_{in}(k)$ at a given scale factor $a_1$.
We also note that radiation pushes the growth of both dark matter and dark energy
perturbations
to later times so that the ratio (which is relevant for $Q$) is basically unchanged.

If a more detailed treatment is desirable, then this can be obtained by following the
discussion in standard texts like e.g.~\cite{Paddy}. We prefer to avoid these additional
complexities here since they are not required for the main points of our work.

In this paper we are working in the Newtonian gauge, but CAMB uses the synchronous
gauge. In order to compare the CAMB output with our formulae, we need to transform
the relevant quantities.
The energy-momentum tensor $T^{\mu}_{\nu}\left(Syn\right)$ in the synchronous gauge is related to the $T^{\mu}_{\nu}\left(Con\right)$ in the conformal Newtonian gauge by the transformation:
\be
T^{\mu}_{\nu}\left(Syn\right)=\frac{\partial\hat{x}^{\mu}}{\partial x^{\sigma}}\frac{\partial x^{\rho}}{\partial\hat{x}^{\nu}}T^{\sigma}_{\rho}\left(Con\right)
\ee
where $\hat{x}^{\mu}$ and $x^{\mu}$ denote the synchronous and the conformal Newtonian coordinates respectively. The relevant transformations then are \cite{mabe}
\bea
\delta\left(Syn\right)&=&\delta\left(Con\right)-\alpha\frac{\dot{\bar{\rho}}}{\bar{\rho}}\\
V\left(Syn\right)&=&V\left(Con\right)-\alpha\left(1+w\right)k^2\\
\delta p\left(Syn\right)&=&\delta p\left(Con\right)-\alpha\dot{\bar{p}}\\
\sigma\left(Syn\right)&=&\sigma\left(Con\right)
\eea
where $\bar{\rho}$ is the energy density at background, $\delta$ is the density contrast, $V$ is the velocity perturbation, $\delta p$ is the pressure perturbation, $\sigma$ is the anisotropic stress and the function $\alpha= \hat{x}^{0}-x^{0}=\left(\dot{h}+\dot{\eta}\right)/2k^2$. Here $\eta$ and $h$ are two scalar fields characterizing the scalar modes of the metric perturbation in synchronous gauge. This transformation also applies to individual species when more than one particle species contributes to the energy-momentum tensor.

\section{Decaying modes}

In the main text we were able to find simple approximate solutions to the full system of
equations
by neglecting certain terms, especially those linking $\delta$ and $V$ for the dark energy
perturbations. This works very well for the parameter values of interest to us,
specifically
$w\le 0$ and $0\le \cs\le 1$. For $\cs<0$ we expect rapid growth of the density
perturbations from a well-known instability. This instability, however, requires exactly
the
coupling which we neglected. In this appendix we want to take a closer look at the
``decaying
modes'' and their behavior especially for $\cs<0$. For this purpose we need to revert to
the
full system. To allow for consistent simplifications we can then cast it in the form of a
single
second-order differential equation by combining Eqs.~(\ref{deltap}) and (\ref{vp}):

\bea
&&\delta''+\left[\frac{3}{a}\left(1-w\right)+\frac{H'}{H}-\frac{A'}{A}\right]\delta'+ \label{gensolut}\\
\nonumber
&&+\left[\frac{3}{a^2}\left(\cs-w\right)\left[\left(2-3\cs\right)+\frac{aH'}{H}-a\frac{A'}{A}\right] + \right.\\
\nonumber
&& \left.+\frac{Ak^2\cs}{\left(a^2H\right)^2}\right]\delta
+\frac{\left(1+w\right)A}{\left(a^2H\right)^2}k^2\phi=0
\eea
where:
\bea
A &=& 1+\frac{9a^2H^2\left(\cs-w\right)}{k^2} \label{eq:A}\\ 
A' &=& \frac{9\left(2aH^2+2a^2HH'\right)\left(\cs-w\right)}{k^2} \label{eq:Adot}.
\eea

We expect to see rapid perturbation growth for imaginary sound speeds on small scales.
For $k\gg 1$ then $A\simeq 1$ and $A' \simeq 0$; Eq. (\ref{gensolut}) becomes:
\bea
&&\delta''+\frac{3}{2a}\left(1-2w\right)\delta'+
\left[\frac{k^2\cs}{H_{0}^{2}\Omega_{m}a}\right.+\\ \nonumber
&& +\left. \frac{3}{2a^2}\left(\cs-w\right)\left(1- 6\cs\right)\right]\delta
+\frac{\left(1+w\right)}{\left(a^2H\right)^2}k^2\phi=0.
\label{gensolut1}
\eea

At the sound horizon $\cs k^2 =(aH)^2 = H_0^2 \Omega_m / a$ so that for sub-sound-horizon
modes
$\nu(a)^2 \equiv k^2\cs a/(H_{0}^{2}\Omega_{m})\gg|(\cs-w)(1- 6\cs)|$, and we can neglect
the contribution from the second term. $\nu$ quantifies how deep the mode is inside
the sound horizon, and it will become complex for $\cs<0$ (notice also that $\nu$ grows as
$\sqrt{a}$). We are then left with:
\be
\delta''+\frac{3}{2a}\left(1-2w\right)\delta'
+ \frac{k^2\cs}{H_{0}^{2}\Omega_{m}a}\delta\\
+\frac{\left(1+w\right)}{\left(a^2H\right)^2}k^2\phi=0
\label{gensolut2}
\ee
The full solution to this equation contains the one found earlier, Eq.~(\ref{eq:dpsh}),
as well
as two additional ones,
\bea
\delta_1 &\propto&  \nu(a)^{-n} J_{n}(2\nu(a)) \Gamma(3/2-3w)\\
\delta_2 &\propto&  \nu(a)^{-n} J_{-n}(2\nu(a)) \Gamma(1/2+3w)
\eea
where $n=(1-6w)/2$. If $\cs<0$ then the absolute
value of the Bessel functions will grow exponentially fast since its argument is complex,
as
expected for the instability.

For super-horizon modes ($k\ll 1$) Eq.~(\ref{eq:A}) reduces to $A
=9a^2H^2\left(\cs-w\right)/k^2$ and Eq.~(\ref{gensolut}) becomes:
\be
\delta''+\frac{5-6w}{2a}\delta'+\frac{9\left(\cs-w\right)}{2a^{2}}\delta
+\frac{9\left(1+w\right)\left(\cs-w\right)}{a^2k^2}k^{2}\phi= 0.
\ee

The full solution to this equation contains the one found earlier, Eq.~(\ref{eq:dsub}),
as well as the following ones,
\bea
\delta_1 &\propto&  a^{\frac{3}{4}\left(-1+2w-\sqrt{-8\cs+(1+2w)^2}\right)}\\
\delta_2 &\propto&  a^{\frac{3}{4}\left(-1+2w+\sqrt{-8\cs+(1+2w)^2}\right)}
\eea

The interesting term here is the square root, in order to find a growing 
solution it needs to be real: $-8\cs+(1+2w)^2\geq0$. This is the case for
\be
(1+2w)^2 \geq 8 \cs .
\ee
This means that if we want to keep the sound speed positive we need to have an
equation of state parameter that is either very negative or well larger than
$-1/2$. In either case the solutions are decaying for $w<0$.

\end{appendix}

\end{document}